\documentclass[aps,showpacs,preprintnumbers,amsmath, amssymb]{revtex4}

\usepackage{float}
\usepackage{graphics,epsfig}
\usepackage{graphicx}
\usepackage{dcolumn}
\usepackage{bm}

\begin{document}
\baselineskip=0.8 cm

\title{{\bf Studies of no light scalar hair behaviors for compact reflecting stars in a box}}
\author{Yan Peng$^{1}$\footnote{yanpengphy@163.com}}
\affiliation{\\$^{1}$ School of Mathematical Sciences, Qufu Normal University, Qufu, Shandong 273165, China}

\vspace*{0.2cm}
\begin{abstract}
\baselineskip=0.6 cm
\begin{center}
{\bf Abstract}
\end{center}

We study condensation of scalar fields around compact reflecting stars in a box.
We propose no light scalar hair behaviors that the neutral compact star
cannot support the existence of small single positive mass scalar field
in it's exterior spacetime for very general self-interacting potentials.
Moreover, we find that this no light scalar hair property can be violated by including
an additional Maxwell field coupled to the scalar field in a charged compact reflecting star background.

\end{abstract}

\maketitle
\newpage
\vspace*{0.2cm}

\section{Introduction}

The no-scalar-hair theorem is a famous physical
characteristic of black holes \cite{Bekenstein,Chase,Ruffini-1}.
It was found that the static massive scalar fields cannot
exist in asymptotically flat black holes, for references see \cite{Hod-1}-\cite{Brihaye}
and a review see \cite{Bekenstein-1}.
This property is usually attributed to the
fact that the horizon of a classical black hole irreversibly absorbs matter and radiation fields.
Along this line, one naturally want to know whether this no scalar hair behavior is a
unique property of black holes. So it is interesting to explore possible similar no scalar hair theorem
in other horizonless curved spacetimes.

Lately, hod found a no-scalar-hair theorem for asymptotically flat horizonless neutral compact reflecting
star with a single massive scalar field and specific types of the potential \cite{Hod-6}.
Bhattacharjee and Sudipta further extended the discussion to spacetimes with a positive cosmological constant \cite{Bhattacharjee}.
In fact, the no scalar hair behavior also exists for massless
scalar field nonminimally coupled to gravity in the  neutral compact reflecting star background \cite{Hod-7}.
On the other side, a simple way to make the scalar field easier to condense is putting the black hole in a box.
It was believed that the box boundary plays a role of the infinity
potential to make the fields bounce back and condense around the black hole.
In fact, it was found that the low frequency scalar field perturbation can trigger
superradiant instability of the charged RN black hole in a box
and the nonlinear dynamical evolution can form a quasi-local hairy black hole \cite{Dolan,Carlos, Supakchai, Nicolas}.
So it is interesting to examine whether there is no-scalar-hair theorem for neutral compact
reflecting star in a box. And it is also interesting to extend the discussion
by including an additional Maxwell field to examine whether no-scalar-hair
theorem works in the charged compact reflecting star background.

The next sections are planed as follows. In section II, we show the no light scalar hair behaviors
in the neutral compact reflecting star in a box.
In section III, we violate this no light scalar hair property by adding a Maxwell field and numerically obtain the
hairy charged compact reflecting star solutions. We will summarize our main results in the last section.

\section{No light scalar hair theorem for neutral compact star}

We firstly consider a spherically symmetric neutral compact reflecting star in a box with a single scalar field
in the background of a four dimensional asymptotically flat gravity.
We fix radial coordinates $r=r_{s}$ as the radius of the compact star and $r=r_{b}$ as the box boundary.
And the corresponding Lagrange density is given by \cite{Hod-6}:
\begin{eqnarray}\label{lagrange-1}
\mathcal{L}=R-\partial_{\alpha} \psi \partial^{\alpha} \psi-V(\psi^{2}),
\end{eqnarray}
where $V(\psi^{2})$ is the self-interacting potential of the scalar field with positive mass or $\dot{V}(0)>0$.
Our discussion includes wide types of potentials with $\dot{V}(\psi^{2})=\frac{dV(\psi^2)}{d\psi^2}$
as an elementary function of $\psi^{2}$, see \cite{Colpi,Friedberg,Ho,Kaup,Lee,Mielke,Ruffini,Schunck}.
For elementary functions, we mean functions constructed with several basic elementary functions
of C, $x^{n}$, $a^{x}$, $Log (x)$, Sin(x) $\cdots$.
In order to see this clearly, we take
a simple form $V(\psi^{2})=a_{1}\psi^{2}+a_{2}\psi^{4}+a_{3}\psi^{6}+\cdot\cdot\cdot+a_{N}\psi^{2N}$
with N as an arbitrary integer as an example. Here, $a_{1}$ can be explained as the 
mass of the scalar field and terms of $\psi^{4}, \psi^{6}, \cdot\cdot\cdot, \psi^{2N}$
correspond to repulsive or attractive effects according to the sign of the coefficients.
In order for the no-scalar-hair theorem in \cite{Hod-6} to work, we need to impose strict conditions that $\dot{V}(\psi^{2})>0$ for all $\psi$.
In contrast, discussion of no light scalar hair behaviors in the following only desires $\dot{V}(0)=a_{1}>0$,
which naturally holds for positive mass scalar field with both repulsive and attractive effects.

Considering the scalar field's backreaction on the metric, we take the
deformed four dimensional compact star solution as \cite{Hod-6}
\begin{eqnarray}\label{AdSBH}
ds^{2}&=&-g(r)h(r)dt^{2}+\frac{dr^{2}}{g(r)}+r^{2}(d\theta^{2}+sin^{2}\theta d\varphi^{2}).
\end{eqnarray}
where the metric solutions satisfies $g(\infty)=1$ and $h(\infty)=1$
in accordance with asymptotically flat behaviors at the infinity.

For simplicity, we study the scalar field with only radial dependence in the form $\psi=\psi(r)$.
From above assumptions, we obtain equations of motion as
\begin{eqnarray}\label{BHg}
\psi''(r)+\frac{g'(r)\psi'(r)}{g(r)}+\frac{h'(r)\psi'(r)}{2h(r)}+\frac{2\psi'(r)}{r}-\frac{\dot{V}}{2g}\psi=0,
\end{eqnarray}
\begin{eqnarray}\label{BHpsi}
\frac{1}{2}\psi'(r)^{2}+\frac{g'(r)}{rg(r)}-\frac{1}{r^2g(r)}+\frac{1}{r^2}+\frac{V(\psi^{2})}{2g}=0,
\end{eqnarray}
\begin{eqnarray}\label{BHpsi}
h'(r)-rh(r)\psi'(r)^2=0.
\end{eqnarray}
Here, we define the expression as $\dot{V}=\frac{d V(\psi^{2})}{d \psi^{2}}$.

In addition, we impose reflecting boundary conditions for the scalar field
at the surface of the compact star and the box boundary. So the scalar field vanishes
at the boundaries as
\begin{eqnarray}\label{InfBH}
&&\psi(r_{s})=0,~~~~~~~~~\psi(r_{b})=0.
\end{eqnarray}

According to the boundary conditions (6), one deduce that the function
of the scalar field must have (at least) one extremum point $r=r_{peak}$
between the surface $r_{s}$ of the reflecting star and the box boundary $r_{b}$.
At this extremum point, the scalar field is characterized by the relations
\begin{eqnarray}\label{InfBH}
\{ \psi'=0~~~~and~~~~\psi \psi''\leqslant0\}~~~~for~~~~r=r_{peak}.
\end{eqnarray}

We have assumed that $\dot{V}=\frac{dV(\psi^{2})}{d\psi^{2}}$ is a
elementary function of $\psi^{2}$, so $\dot{V}$ is continuous as a function of $\psi^{2}$.
And $\dot{V}$ is locally positive according to properties of continuity.
For each given form of the potential, there is a fixed constant $\delta>0$ satisfying $\dot{V}(\psi^{2})>0$
with $||\psi||=max\{|\psi(r)|~|r_{s}\leqslant r \leqslant r_{b}\}<\sqrt{\delta}$.
Then, we arrive at the inequality
\begin{eqnarray}\label{BHg}
\psi''(r)+[\frac{2}{r}+\frac{g'(r)}{g(r)}+\frac{h'(r)}{2h(r)}]\psi'(r)-\frac{\dot{V}}{2g}\psi<0~~~for~~~r=r_{peak}~~and~~||\psi||<\sqrt{\delta},
\end{eqnarray}
in contradiction with the equation (3).
Finally, we arrive at the conclusion that small scalar hair of $||\psi||<\sqrt{\delta}$ cannot
exist in the neutral compact reflecting star background.

\section{Charged scalar hairy compact star in a box}

In this part, we study the model constructed by a charged scalar field coupled to a Maxwell field in
the background of four dimensional asymptotically flat charged compact reflecting star in a box.
And the general Lagrange density reads \cite{Pallab Basu}:
\begin{eqnarray}\label{lagrange-1}
\mathcal{L}=R-F^{MN}F_{MN}-|\nabla_{\mu} \psi-q A_{\mu}\psi|^{2}-m^{2}\psi^{2},
\end{eqnarray}
where q and $m$ are the charge and mass of the scalar field $\psi(r)$ respectively.
And $A_{\mu}$ stands for the ordinary Maxwell field.

For simplicity, we study matter fields with only radial dependence in the form
\begin{eqnarray}\label{symmetryBH}
A=\phi(r)dt,~~~~~~~~\psi=\psi(r).
\end{eqnarray}

With above assumptions and the metric (2), we obtain equations of motion as
\begin{eqnarray}\label{BHg}
\psi''(r)+\frac{g'(r)\psi'(r)}{g(r)}+\frac{h'(r)\psi'(r)}{2h(r)}+\frac{2\psi'(r)}{r}+\frac{q^2\psi(r)^2\phi(r)^2}{g(r)^2h(r)}-\frac{m^2}{g}\psi=0,
\end{eqnarray}
\begin{eqnarray}\label{BHphi}
\phi''+\frac{2\phi'(r)}{r}-\frac{h'(r)\phi'(r)}{2h(r)}-\frac{q^2\psi(r)^2\phi(r)}{2g(r)}=0,
\end{eqnarray}
\begin{eqnarray}\label{BHpsi}
\frac{1}{2}\psi'(r)^{2}+\frac{g'(r)}{rg(r)}+\frac{q^2\psi(r)^2\phi(r)^2}{2g(r)^2h(r)}+\frac{\phi'(r)^2}{g(r)h(r)}-\frac{1}{r^2g(r)}+\frac{1}{r^2}+\frac{m^2}{2g}\psi^2=0,
\end{eqnarray}
\begin{eqnarray}\label{BHpsi}
h'(r)-rh(r)\psi'(r)^2-\frac{q^2r\psi(r)^2\phi(r)^2}{g(r)^2}=0,
\end{eqnarray}

These equations are nonlinear and coupled, so we use the
shooting method to integrate the equations from the compact star
surface $r_{s}$ to the box boundary $r_{b}$ to search for numerical
solutions. Around $r_{s}$, the solutions can be expanded as

\begin{eqnarray}\label{InfBH}
&&\psi(r)=aa (r-r_{s})+bb (r-r_{s})^2+\cdots,\nonumber\\
&&\phi(r)=aaa+bbb (r-r_{s})+\cdots,\nonumber\\
&&g(r)=1+AA (r-r_{s})+\cdots,\nonumber\\
&&h(r)=AAA+BBB (r-r_{s})^2+\cdots,
\end{eqnarray}
where the dots denote higher order terms.
Putting these expansions into equations of motion and considering leading
terms, we have three independent parameters $aa$, $aaa$ and $AAA$ left to describe the solutions.
With the rescaling $r\rightarrow \alpha r$, we could set $r_{b}=1$. Near the box boundary $(r=1)$, the asymptotic
behaviors of the matter fields are $\psi(1)=0$.
In addition, we can also take $h(1)=1$ with the symmetry
$h\rightarrow \beta^2h,~\phi\rightarrow \beta\phi,~t\rightarrow\frac{t}{\beta}$.

Now, we show the numerical hairy compact star solutions
with $q=1$, $m^{2}=1$ and $aa=\frac{1}{100}$ in Fig. 1.
In the left panel, the scalar field satisfies the reflecting
conditions that $\psi(r_{s})=0$ and $\psi(r_{b})=0$.
In the middle panel, the vector field
increases as a function of the radial coordinate. We also show behaviors of the
metric solutions $h(r)$ in the right panel.
When neglecting the matter fields'
backreaction on the metric, we will have $h(r)=1$ and in contrast,
the curves in the right panel show that the metric is deformed by backreaction of matter fields.
So we find that the no-scalar-hair theorem in neutral compact reflecting star usually
doesn't exist in charged compact reflecting star.

\begin{figure}[h]
\includegraphics[width=155pt]{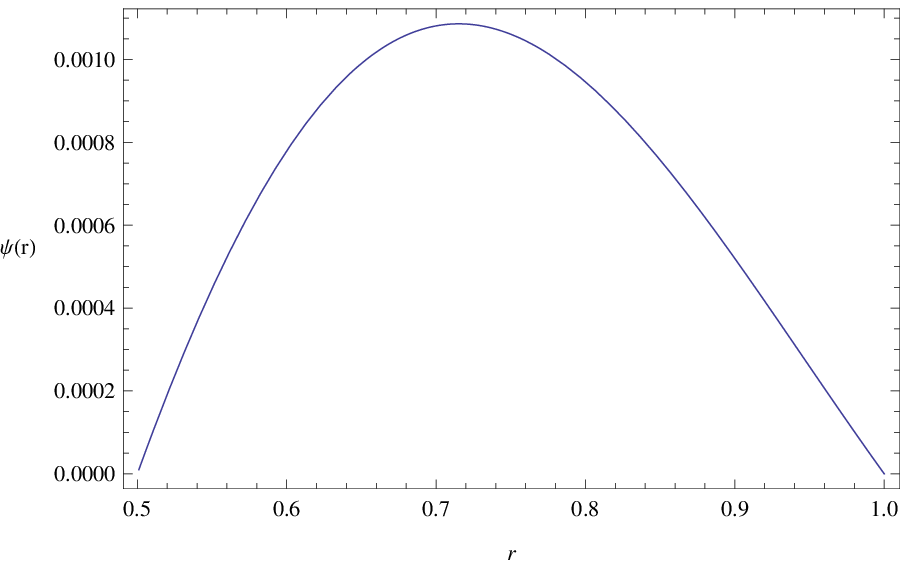}\
\includegraphics[width=155pt]{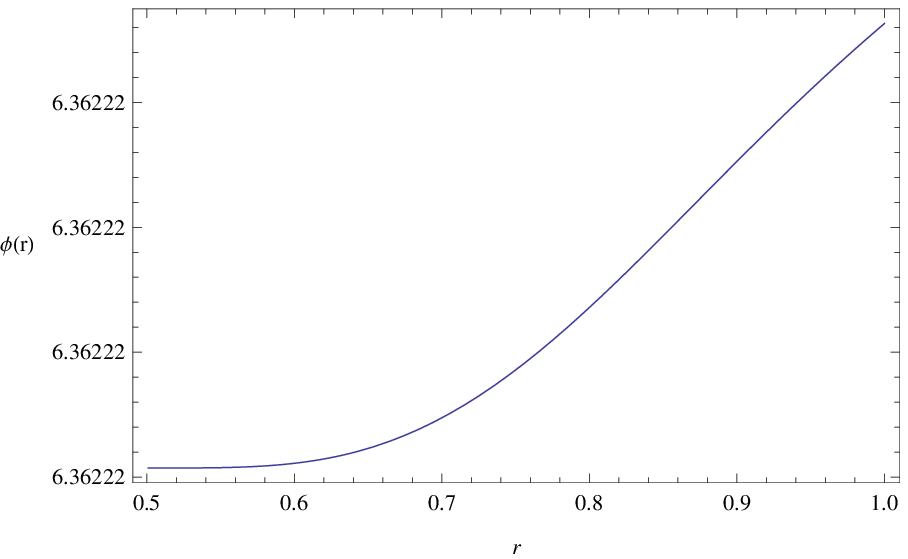}\
\includegraphics[width=155pt]{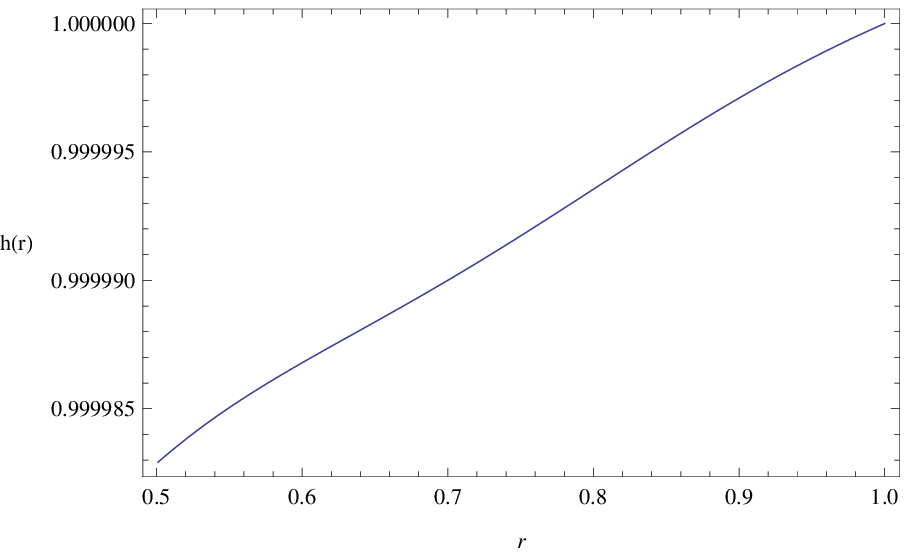}\
\caption{\label{EEntropySoliton} (Color online) We plot solutions as a function of the
radial coordinate $r$ with $q=1$, $m^{2}=1$ and $aa=\frac{1}{100}$. The left panel shows behaviors
of $\psi(r)$, the middle panel corresponds to the vector field $\phi(r)$
and the right panel represents the values of $h(r)$.}
\end{figure}

We also plot the scalar field in cases with $q=1$, $m^{2}=1$ and various $aa$ in Fig. 2.
It can be easily seen from Fig. 2 that when we choose a fixed value for $aa$ as $aa=aa_{1}=\frac{1}{100}$, we obtain
a maximum value for the scalar field as $||\psi||_{1}\thickapprox0.001085$.
As we choose a smaller value of $aa=aa_{2}=\frac{1}{1000}=\frac{1}{10}aa_{1}$,
we have $||\psi||_{2}\thickapprox0.0001085\thickapprox\frac{1}{10}||\psi||_{1}$.
This is reasonable since small scalar field can be seen as a perturbation
to the charged compact star background and equation (11) can be treated
as a linear equation of $\psi$. So we can obtain arbitrarily small $||\psi||$
by continuing to choose smaller parameter $aa$.
Finally, we conclude that the no small scalar hair property is violated
in this charged compact star background.

\begin{figure}[h]
\includegraphics[width=200pt]{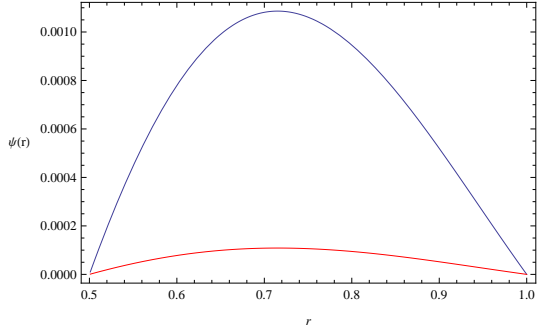}\
\caption{\label{EEntropySoliton} (Color online) We show behaviors of the scalar field
with $q=1$, $m^{2}=1$ and various $aa$ from top to bottom as $aa=aa_{1}=\frac{1}{100}$ (blue) and
$aa=aa_{2}=\frac{1}{1000}$ (red).}
\end{figure}

\section{Conclusions}

We studied condensation of scalar fields around compact reflecting stars enclosed in a box.
We showed no light scalar hair behaviors that the neutral compact star cannot support the existence of small scalar field
in it's exterior spacetime for positive mass scalar fields with a very general self-interacting potential.
This no light hair property may imply that it is not easy for the positive mass
scalar field to condense in the neutral compact reflecting star background.
In contrast, we numerically obtained scalar hairy charged compact reflecting star solutions
and further demonstrated that the no light scalar hair properties are violated in the charged compact reflecting star
when including an additional Maxwell field coupled to the scalar field.
It is interesting to study the stability of hairy charged compact reflecting stars under kinds of perturbations and we plan to
carry out this research in the future work.

\begin{acknowledgments}

This work was supported by the National Natural Science Foundation of China under Grant No. 11305097;
the Shaanxi Province Science and Technology Department Foundation of China
under Grant No. 2016JQ1039.

\end{acknowledgments}

\end{document}